\documentclass{mn2e}
\usepackage{times}
\usepackage{epsf}

\newcommand{\etal}{{et al}\/.}
\newcommand{\rosat}{\textit{ROSAT}}
\newcommand{\einstein}{\textit{Einstein}}
\newcommand{\chandra}{\textit{Chandra}}
\newcommand{\vlba}{\textit{VLBA}}
\newcommand{\vla}{\textit{VLA}}
\newcommand{\hst}{\textit{HST}}

\newcommand{\iso}{\textit{ISO}}
\newcommand{\css}{3C\,48}

\begin{document}

\title[The X-ray and radio components in \css]
{The relationship between the X-ray and radio components in the
Compact Steep Spectrum Quasar \css}

\author[D.M. Worrall \etal]{D.M.\ Worrall$^{1,2}$, 
M.J.\ Hardcastle$^1$, T.J.\ Pearson$^3$ and A.C.S.\ Readhead$^3$\\
$^1$ Department of Physics, University of Bristol, Tyndall Avenue,
Bristol BS8 1TL\\
$^2$ Harvard-Smithsonian Center for Astrophysics, 
60 Garden Street, Cambridge, MA 02138, USA\\
$^3$ California Institute of Technology,  Mail Stop 105-24, 
Pasadena, CA 91125, USA\\}

\maketitle

\label{firstpage}

\begin{abstract}
We combine results from \rosat, \chandra, and multifrequency \vlba\
observations of the Compact Steep Spectrum quasar \css\ in an attempt
to understand why the radio source is so small and unusual.  The X-ray
spectrum shows no evidence for the excess absorption which might have
allowed us to conclude that \css\ is small because it is bottled up
by cold neutral gas.  We infer that the X-ray spectrum of the nucleus
is made up of a soft, variable, steep-spectrum component, and
harder, power-law emission of slope consistent with the 1~GHz radio
spectrum.  The similarity of \css's X-ray to radio ratio to that
seen in core-dominated radio-loud quasars leads us to examine the
possibility that the harder X-ray emission is inverse Compton
radiation from the radio source, which is more than 99 per cent
resolved in our \vlba\ data.  The weak ($3\sigma$) evidence that we
find for a proper motion of $0.5\pm0.2c$ in a compact radio component
about 0.05 arcsec from the core implies that if this component has a
highly relativistic bulk motion, it is at a very small angle to the
line of sight.  However, stringent requirements on the jet opening angle
make it unlikely that all the X-ray emission is from a
fast jet which sees boosted Cosmic Microwave Background emission and
emits beamed X-rays in the observer's frame.  If the unusual radio
structures are intrinsically one-sided and unbeamed, the inverse
Compton mechanism can provide an appreciable fraction of the X-ray
emission if the magnetic field strength is a factor of six to ten
below that which gives equal energy in radiating relativistic
particles and magnetic fields and roughly minimizes the total energy
in the source.  It remains possible that the unresolved X-ray
emission arises from close to the central engine, either as an
embedded blazar or associated with the accretion processes.
\end{abstract}

\begin{keywords}
quasars: individual: \css --- radiation mechanisms: non-thermal --- radio
continuum: galaxies --- X-rays: galaxies
\end{keywords}

\section{Introduction}
\label{sec:intro}

Most radio-loud quasars contain a compact, parsec-scale, flat-spectrum
radio core, and lower-surface-brightness extended kiloparsec lobes.
The wide range in the ratio of core to total lobe emission is usually
attributed to Doppler boosting of the core in those sources that
contain a jet aligned close to the observer's line of sight.  But a
substantial fraction of radio-loud quasars do not fit into this
`unified model'. These are the {\it compact steep-spectrum quasars}
(CSS quasars), which are characterized by a steep power-law radio
spectrum, usually with a low-frequency turnover below 1~GHz due to
synchrotron self absorption, and sometimes with spectral flattening at
high frequencies.  The radio structures extend no more than about 15
kpc. About 30 per cent of radio sources found in high-frequency
surveys have spectra of this type (Peacock \& Wall 1982).  A review is
found in O'Dea (1998).

The essential difference between CSS quasars and those with flat core
spectra has not yet been identified.  The CSS sources, as a class,
cannot be normal extended double sources seen end-on, because in most
cases the overall size is smaller than the lobe-widths of normal
double sources, and the low-frequency spectral turnover occurs at
higher frequencies.  There are also too many CSS sources for them all
to be attributed to projection effects (Fanti \etal\ 1990).

The role of relativistic boosting in the VLBI structures may not be as
significant in CSS quasars as in their flat-spectrum counterparts
(O'Dea 1998).  However, in 3C\,147 and 3C\,309.1, for example, the
X-ray flux densities from {\einstein} are smaller than the
inverse-Compton emission predicted from the radio flux densities, thus
providing evidence for bulk relativistic motion (Simon \etal\ 1983;
Kus \etal\ 1990).  Superluminal motion is detected in 3C\,216,
3C\,309.1, 3C\,380 and 3C\,138 (Barthel \etal\ 1988; Kus \etal\ 1993;
Cotton \etal\ 1997), although it is now thought that 3C\,216 and
3C\,380 may be large doubles in projection rather than true members of
the CSS class (O'Dea 1998).  Less controversially, in contrast to
flat-spectrum quasars it appears that the jets of CSS quasars have
more complex and distorted morphology and have not escaped from the
host galaxy.  While the underlying cause (whether they are young or
are smothered by dense interstellar gas) remains an issue of debate,
some level of interaction between the radio jets and the local
galactic environment would seem inevitable.

\begin{figure}
\epsfxsize 8.0cm
\epsfbox{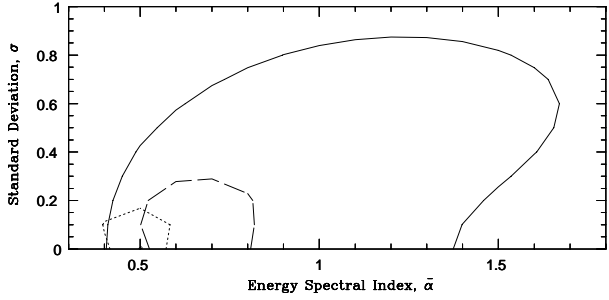}
\caption{90 per cent confidence contours of mean spectral index versus
standard deviation for objects observed with the \einstein\
Imaging Proportional Counter (IPC).  {\it Dotted line\/}: 11
flat-radio-spectrum (FRS) core dominated quasars with mean net IPC
counts = 1316.  {\it Dashed line\/}: 7 steep-radio-spectrum (SRS)
lobe-dominated quasars with mean net IPC counts = 1163.  {\it Solid
line\/}: 7 compact steep-radio-spectrum (CSS) quasars with mean net
IPC counts = 208.  The maximum-likelihood analysis to derive their
class properties is from Worrall (1989).  The {\einstein}-observed CSS
sources are \css, 3C\,147, 3C\,191, 3C\,286, 3C\,287, 3C\,380, and
3C\,309.1. }
\label{fig:xml}
\end{figure}

X-ray measurements of CSS quasars are important for the following
reasons: (a) in conjunction with radio VLBI observations, they place
lower limits on the bulk flow velocity in the emitting regions by the
inverse-Compton argument; (b) they may show evidence for higher
neutral gas density in these objects than in other quasars; (c) they
may be able to show whether an additional emission mechanism (e.g.,
thermal) is operating, or whether all the radiation is consistent with
a nuclear and jet origin; (d) large spectral differences between
objects may indicate that they do not form a homogeneous class.  Our
best composite X-ray knowledge to date comes from objects drawn from the 3CRR
sample (Laing, Riley \& Longair 1983).  \einstein\ observed several
such CSS quasars, but the derived overall knowledge of their soft
X-ray spectra is poor compared with other classes of radio-loud
quasars (Figure~\ref{fig:xml}).  Radio-loud, flat-spectrum
core-dominated quasars have spectra consistent with a power law with a
typical energy spectral index\footnote{Throughout this paper we define
spectral index $\alpha$ in the
sense $f_\nu \propto \nu^{-\alpha}$}
$\alpha \approx 0.5$ (Worrall \& Wilkes 1990), and there is a tendency
for the slope to steepen with decreasing core dominance (Shastri
\etal\ 1993).  It is not known whether the large contour for CSS
sources in Figure~\ref{fig:xml} is because of poor statistics or
because they show a much larger spread in their X-ray spectral
indices.

In this paper we report \rosat\ and \chandra\ X-ray results for
\css, one of the strongest and most peculiar CSS quasars.
We present previously unreported variability between the two
observations with the \rosat\ Position Sensitive Proportional Counter
(PSPC) and within the observation with the High Resolution Imager
(HRI).  In comparison with an earlier \einstein\ Observatory
observation we infer the presence of a two-component spectrum which is
supported by the new \chandra\ data.  We also present new
multifrequency \vlba\ observations which we use to predict the
non-thermal X-ray emission for comparison with the X-ray results.

To ease comparison with earlier work we use a cosmology in which $H_0
= 50$~km s$^{-1}$ Mpc$^{-1}$, $q_0 = 0$.  At \css's redshift of
0.367, 1~arcsec corresponds to 6.8~kpc in the source frame.

\subsection{Properties of \css}
\label{sec:c48prop}

\css\ was the second quasar to be recognized (Matthews \& Sandage
1963; Greenstein \& Matthews 1963).  It is classified as a CSS source
on the basis of its steep radio spectrum and its small angular
size. The integrated radio spectrum shows a low-frequency peak at
about 100~MHz and is reasonably straight (a slight concave curvature)
from $\sim 500$~MHz to $>20$~GHz with $\alpha \approx 0.75$ at 1~GHz
(Ott \etal\ 1994).  Most of the radio emission occurs on a scale less
than about 0.6~arcsec ($< 4$ kpc), although at low frequencies there
is emission on scales up to 3.5 arcsec (Hartas \etal\ 1983; Woan
1992). VLBI images [Wilkinson \etal\ 1990, 1991 (1.6~GHz); Simon
\etal\ 1990 (329~MHz); Nan \etal\ 1991 (608~MHz)] show a compact,
inverted-spectrum `core', A, at the southern end of the source, with a
`hotspot', B, 50~mas to the north of A, from which a one-sided, broad
disrupted jet or plume extends $\sim 1$ arcsec to the north and
northeast. This is unlike most other CSS sources which generally show
more-or-less symmetrical double or triple structures. The VLBI images
show no sign of a counterjet, but a high dynamic-range \vla\ image at
8 GHz (Briggs 1995) finds a component on the counterjet side, with an
inferred jet-counterjet ratio around 100:1. There have been no VLBI
images published at frequencies higher than 1.6~GHz.  The flux density
appears to be variable by a few per cent over decades at frequencies
above a GHz (Ott \etal\ 1994).

\css\ has associated optical nebulosity. Broad and narrow [OIII] and
[OII] emission-line regions form a chain of compact knots which
extends to 6 arcsec north of the core, beyond the radio structures;
photoionization by the quasar nucleus is believed to play the dominant
role in this gas excitation (Axon \etal\ 2000).  Excess continuum and
emission-line nebulosity closer to the quasar core, and in particular
at about 1 arcsec northeast of the nucleus, is variously interpreted
as the nucleus of a galaxy merging with the host galaxy of \css\
(Stockton \& Ridgway 1991) or a radio-jet-related star-forming region
(Chatzichristou, Vanderriest \& Jaffe 1999).  \css\ has unusually
strong infrared emission (Neugebauer, Soifer, \& Miley 1985),
comparable with the most luminous infrared galaxies, and the detection
of CO emission (Scoville \etal\ 1993) implies the presence of about
twice as much molecular gas as in ultraluminous {\it IRAS\/} galaxies.
This evidence, together with the irregular appearance of the galaxy in
optical continuum (Stockton \& Ridgway 1991), supports the idea of a
recent merger.  Fabian \etal\ (1987) suggest that the extensive
emission-line nebulosity is photoionized gas resulting from a cooling
flow of $\sim 100 M_{\sun}$~yr$^{-1}$ rather than a galaxy merger.

\css\ was detected in X-rays with the \einstein\ IPC on 1980 Jul 25
(Wilkes \& Elvis 1987). The spectrum was consistent with a power-law
with $\alpha = 0.7\,^{+0.6}_{-0.4}$ 
and a flux density at 1~keV of $f = 0.50\,^{+0.14}_{-0.10}$ $\mu$Jy (90
per cent errors for 2 interesting parameters; $\Delta \chi^2 +
4.6$). (Our reanalysis of the data shows the quoted error in flux
density is more appropriate for $1\sigma$ for 2 interesting
parameters; $\Delta \chi^2 + 2.3$.)

\section{Observations and Analysis Methods}

\subsection{\rosat}
\label{sec:rosobs}

The \rosat\ observations, whose dates and exposure times are given in
Table~\ref{tab:xobs}, were made with the source in the center of the
telescope field of view and in normal `wobble' mode during the pointed
phase of the mission.  \css\ was observed twice with the PSPC and
once with the HRI. The data were corrected for instrumental effects
and the motion of the satellite.  Our analysis of the HRI data
includes `dewobbling' (Harris \etal\ 1998) to remove some of the
time-dependent aspect uncertainties arising from pixel-to-pixel gain
variations in the aspect camera CCD, as well as the use of a broadened
PSF (calculated by fitting to data within 10 arcsec of the source
centroid) to take account of residual aspect uncertainties.  Our
analysis used the Post Reduction Off-line Software ({\sc PROS} X-ray package
which runs in the {\sc IRAF} environment.

\begin{table*}
\caption{\rosat\ and \chandra\ observations and single-component power-law spectral fits}
\begin{tabular}{llllrlllll}
\hline
Source & z & Date & Inst. & $T_{\rm Exp}$ & Net & $\alpha$ & $N_H$ &
1-keV flux & $\chi^2$/dof\\
& & &  &  (s) & Counts &  & $10^{20}$ cm$^{-2}$& density ($\mu$Jy) & \\
& & &  &  &  &  &  & &\\
\hline
\css\ & 0.367 & 1992 Jan 26-27&\rosat\ PSPC&3,442& $1240\pm38$& $1.52\pm0.20$&
$3.56\pm 0.61$& $0.77\pm 0.07$& 
30.8/26\\
& & 1992 Aug 9&\rosat\ PSPC& 2,250& \phantom{1}$566\pm25$& $1.51\pm0.32$&
$3.64^{+0.96}_{-0.91}$& $0.53^{+0.08}_{-0.07}$& 35.6/25\\
& & 1994 Aug 11-13&\rosat\ HRI&37,020&$6000 \pm 166$&--&--&$1.15 \pm 0.03$&--\\
& & 2002 Mar 6&\chandra\ ACIS-S&9,225&$2522 \pm 50^a$&
$1.35 \pm 0.08$&$4.35^b$&$0.80 \pm 0.03$&81.1/64\\
\hline
\end{tabular}
\label{tab:xobs}
\begin{minipage}{18cm}
a. In the annulus of the wings of the PSF used for spectral fitting.\\
b. Fixed to the radio-derived value of
$N_H$ for our Galaxy of $4.35 \times 10^{20}$ cm$^{-2}$ (Elvis,
Lockman \& Wilkes 1989).\\
Errors in the spectral parameters are $1\sigma$ for 2 interesting
parameters ($\chi^2 + 2.3$).  Flux density is before
absorption.  In converting the HRI count rate to a flux density,
the spectral parameters derived from the PSPC data have
been used, and the quoted error on flux density reflects the
statistical error on the count rate only.  
Prieto (1996) fitted the January 1992
spectrum, as part of a study of 3C radio sources, and reports
parameter values consistent with ours.
\end{minipage}
\end{table*}

Spectral distributions were extracted for each PSPC observation using
an on-source circle of 3 arcmin radius and local background from an
annulus of radii 3 and 5.7 arcmin. The large on-source region relative
to most of the source counts was chosen so as to avoid losing soft
photons because of the `Ghost Image' effect of the PSPC (Briel \etal\
1994).  Radial profiles were fitted to models convolved with the Point
Spread Function (PSF) using home-grown software whose algorithms are
described in Birkinshaw (1994). This software takes into account any
contribution of the model in the background region.  A fit was first
made to a one-component point-source model.  We then examined the
reduction in $\chi^2$ when adding in a $\beta$ model of surface
brightness $b_{\rm x}(\theta) \propto (1 + [\theta/\theta_{\rm
c}]^2)^{-3\beta + 0.5}$, used to represent the presence of X-ray
emitting gas in hydrostatic equilibrium.

The presence of an extended X-ray atmosphere in the HRI
observation of \css\ has been reported previously (Crawford \etal\
1999; Hardcastle \& Worrall 1999).
This is re-examined in the context of the new \chandra\ observations
in \S\ref{sec:xspat}.
Our analysis of the \rosat\ data and the comparison with earlier
\einstein\ data was completed before
the \chandra\ data became available in the public archive, and so was
not guided by information from the higher-quality data set.

\subsection{\chandra}
\label{sec:chandraobs}

\css\ was observed with the back-illuminated CCD chip, S3, of the
Advanced CCD Imaging Spectrometer (ACIS) on board \chandra\ on 2002
March 6.  
The observation was part of a study of a small sample of CSS and
Giga-Hertz Peaked Spectrum sources by Siemiginowska \etal\ (2003a).
Details of the instrument and its modes of operation can be
found in the \chandra\ Proposers' Observatory Guide, available from
http://cxc.harvard.edu/proposer.  The observation was made in the
VFAINT data mode, with a 128-row subarray (giving a 1 by 8 arcmin
field of view) to reduce the CCD readout time to 0.44~s and mitigate
the effects of pileup on the spectrum of the quasar core.  We have
analyzed the archival data set using {\sc CIAO v2.3} and the {\sc
CALDB v2.21} calibration database, and following software `threads'
available from \chandra\ X-ray Center (http://cxc.harvard.edu/ciao).
Events with grades 0,2,3,4,6 are used.  All X-ray spectra are binned
to a minimum of 30 counts per bin, and the time-dependent decline in
quantum efficiency is taken into account using the recommended
methods. There were no intervals of flaring background, and 
the calibrated data have an observation duration of 9.225~ks.

The counts extracted from the circle of radius 2.5 arcsec centred on
the nucleus give 0.35 counts/readout time, for which a pileup fraction
of about 8 per cent is estimated by the {\sc PIMMS} software.  A
similar fraction of piled-up events is deduced by examining the events
with significant signal in the outer pixels of the 5 by 5 event
islands read out in VFAINT data mode.  The effect of pileup is to
combine the energy of multiple soft X-rays in the core of the PSF into
a single harder X-ray and thus flatten the spectrum.  We therefore
used two approaches to extract and model the X-ray spectrum.  The
first approach follows Gambill \etal\ (2003) in extracting the counts
from an annular region of inner radius 0.5 arcsec and outer radius 2.5
arcsec, to exclude the piled up core.  The resulting fluxes were then
corrected for missing flux using a spectrally weighted PSF extracted
using the {\sc ChaRT} and {\sc MARX} software (to model the mirror and
instrument responses) and smoothed by a small fraction of an arcsec to
give a good match to the radial profile of the data within 1 arcsec of
the nucleus (thus taking into account residual effects of aspect
uncertainties and pixelization).  The second approach was to fit data
extracted from a 2.5 arcsec radius circle either ignoring the effects
of pileup or attempting to correct for it using the method of Davis
(2001) as implemented in {\sc Sherpa}.  We found
the pileup model overestimated the amount of pileup and steepened and
increased the intensity of
the spectrum so much as to be inconsistent with results from the annular
extraction. We attribute this to the implementation of the pileup model
running into difficulties for a spectrum with as
much concave curvature as that we measure (see \S\ref{sec:chandraspectrum}).
The spectral indices from our fits without the pileup model
to the data from the 2.5 arcsec radius circle were consistent with
those found from the annulus and had smaller statistical
uncertainties.  However, because of systematic errors introduced by
pileup, the model results and uncertainties reported in
\S\ref{sec:chandraspectrum} are derived from fitting to the data from
the annular extraction region, as obtained using {\sc Sherpa}.
Background for the nuclear spectrum was measured from a source-centred
annulus of radii 6 and 20 arcsec. Our spectral model fitting was over
energies between 0.3 and 7 keV. We used the PSF to search for extended
emission using the method described in \S\ref{sec:rosobs}.

\subsection{\vlba}
\label{sec:vlbaobs}

We observed \css\ with the \vlba\ (including a single \vla\ antenna)
at 1.5, 5.0, 8.4 and 15.4 GHz on 1996 Jan 20. The multi-frequency
approach was chosen to allow us to constrain the spectra of the
compact radio components. The 12 hours of observing time was evenly
divided among the four frequencies.

The data were reduced in the standard manner using the {\sc AIPS}
software package. Pulse calibration data and observations of 3C\,345
were used to perform a first calibration of the phases. We then used a
point model to perform global fringe-fitting on the data, generated
best maps from this process and then, because of the complex structure
of the source, fringe-fitted again, using a model based on the
first-generation maps, to produce the data used in the final
analysis. This process was successful for all the data except for that
from the 15-GHz observations, where a point model for the first
iteration of fringe fitting produced unacceptably poor results. In
this case we used a model based on the 8.4-GHz data for the first
iteration of fringe fitting.

The final datasets were averaged to an integration time of 30 s and
mapped using the {\sc AIPS} task {\sc IMAGR} with natural weighting of
the visibilities. The different observing frequencies (four in the
case of the 1.5-GHz data, two for the others) were averaged in order
to make the maps, with the flux scales being corrected on the
assumption of a spectral index of 0.7. The resulting maps were used in
a number of iterations of antenna-based phase self-calibration until
phase artifacts were largely eliminated. Amplitude self-calibration
was not applied, since this might have rendered the flux scale of the
maps unreliable.

\section{X-ray spectrum}
\label{sec:xspec}

\subsection{\rosat}
\label{sec:rosspectrum}

The spectrum from each PSPC observation gives an acceptable fit to a
single-component power law with free $\alpha$, $N_H$, and
normalization (Table~\ref{tab:xobs}).  The absorption is slightly low
as compared with the radio-inferred value for the line of sight
through our Galaxy (Elvis \etal\ 1989), although it is consistent within
$\sim 1\sigma$ uncertainties. The spectrum is steep, and there is no
evidence of excess absorption as might be expected were the X-ray
emission seen through the hypothetical dense cold gas that helps to keep the
radio source small.  A single-temperature Raymond-Smith thermal model
fits the data poorly: e.g., the best such thermal fit to the stronger
detection gives $\chi^2_{\rm min}$ = 56 for 26 degrees of freedom.
The addition of a thermal component does not improve the
single-component power-law fits.

\begin{figure}
\begin{center}
\epsfxsize 5cm
\epsfbox{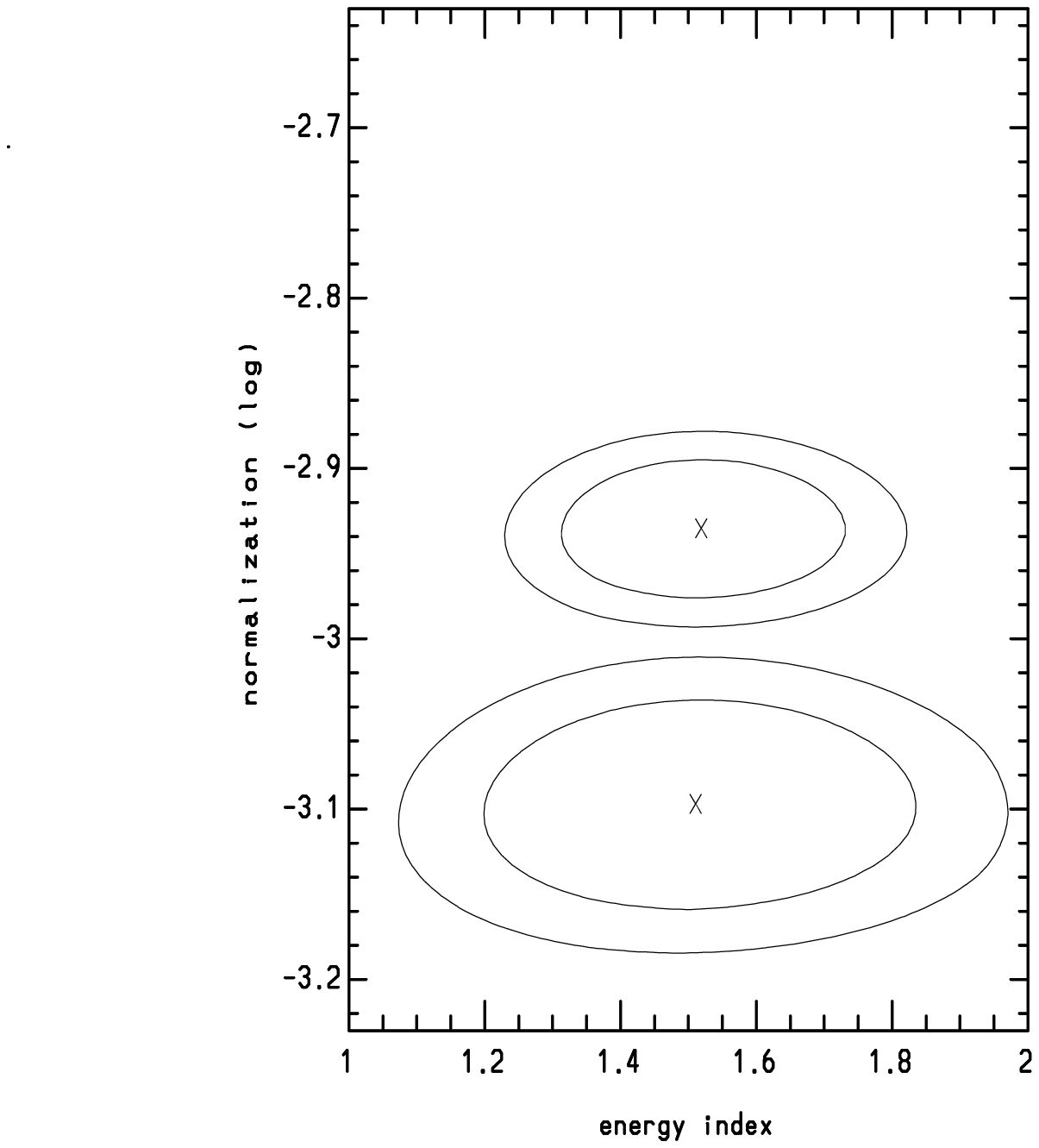}
\end{center}
\caption{Although the X-ray spectral indices in 1992 January and
August are consistent with one another, the normalization was higher
for the earlier observation.  For each observation, the plot shows
contours of equal $\chi^2$ at values of 2.3, and 4.6 above the
minimum, corresponding to uncertainties of 68 and 90 per cent
for two interesting parameters, respectively, and 87 and 97 per cent for
one interesting parameter, respectively.  The normalization, plotted
logarithmically, is in units of keV cm$^{-2}$ s$^{-1}$ keV$^{-1}$ at 1
keV, which, when multiplied by 662.6 gives $\mu$Jy. The upper
contours are for January. 
}
\label{fig:specnorm}
\end{figure}

Although the spectral shape remained remarkably constant between the
observations, the normalization did not, and the flux of the first
observation is $\sim 1.5$ times that of the second
(Table~\ref{tab:xobs} and Figure~\ref{fig:specnorm}).  The \rosat\
data give a significantly steeper power-law spectrum than the earlier
\einstein\ measurement.  The X-ray emission could arise from a power
law which varies both in spectral shape and normalization over time,
but, since the spectral index remained constant between the two
\rosat\ observations, we investigate the possibility that the
\einstein\ and \rosat\ bands are primarily sampling separate
components of emission.  As \einstein\ is sensitive to higher energies
than \rosat\ (although the bands overlap considerably), we model the
spectrum with a function which flattens to higher energies.  Such a
concave curvature might explain why the \rosat\ data slightly prefer
an absorption less than the column density through our Galaxy.  We
fitted each of the three observations (one \einstein\ and two \rosat)
to a two-component power-law model to investigate whether or not it is
possible to find two power-law slopes which, with normalizations as
the only free parameters, fit each observation at least as well as the
single power law.  For this exercise we fixed the absorption to the
Galactic value.  We stepped through values for one power-law index,
fitting the other index and the normalizations, to find a case where
the second power-law index was similar for all three observations.  We
then fixed both power laws and computed $\chi^2$ and the two
normalizations for a fit to each data set.  Table~\ref{tab:xspec}
gives example fits for two components.  Due to the poorly constrained
nature of the fitting we do not claim to have found the best two
slopes.  However, this does show that it is possible that \rosat\ and
\einstein\ are both sampling a power law of slope similar to that of
the radio spectrum, plus a steeper soft excess, with the normalization
of the soft component (and possibly the hard component) varying over
time.

\begin{table}
\caption{Single and 2-component power-law spectral fits with $N_H$ fixed at
the Galactic value}
\begin{tabular}{llllll}
\hline
Observation & $\alpha_{\rm h}$ & $f_{\rm h}$ & $\alpha_{\rm s}$ & 
$f_{\rm s}$ & $\chi^2$/dof\\
 &  &  $\mu$Jy &  &  $\mu$Jy & \\
\hline
\einstein\ 1980& 1.1 & 0.62 & -- & -- & 10.0/7\\
& 1.0 & 0.56 & 2.2 & 0.04 & 9.5/7\\
& 0.9 & 0.52 & 2.2 & 0.07 & 9.2/7\\
& 0.8 & 0.47 & 2.1 & 0.11 & 9.0/7\\
PSPC 1992 Jan& 1.8  & 0.78 & -- & -- & 34.3/27\\
& 1.0 & 0.41 & 2.2 & 0.36 & 31.3/27\\
& 0.9 & 0.39 & 2.2 & 0.38 & 31.3/27\\
& 0.8 & 0.32 & 2.1 & 0.45 & 31.5/27\\
PSPC 1992 Aug& 1.7 & 0.54 & -- & -- & 36.7/26 \\
& 1.0 & 0.30 & 2.2 & 0.23 & 36.0/26 \\
& 0.9 & 0.28 & 2.2 & 0.24 & 36.1/26\\
& 0.8 & 0.23 & 2.1 & 0.29 & 36.1/26\\
\chandra\ 2002& $0.5^{+0.3}_{-0.5}$ & $0.34\pm 0.2$ & $2.2^{+0.7}_{-0.4}$ &
$0.4\pm 0.2$ & 41.9/62 \\
\hline
\end{tabular}
\label{tab:xspec}
\begin{minipage}{8cm}
Only the \chandra\ fit is truly for minimum $\chi^2$: for \einstein\
and \rosat\ example fits are shown.
$f_{\rm h}$ and $f_{\rm s}$ are the 1~keV flux densities.
$N_H$ is fixed at the Galactic value of $4.35 \times 10^{20}$
cm$^{-2}$ throughout this table.  The power laws cross at an energy of
$(f_{\rm s}/f_{\rm h})^{1/(\alpha_{\rm s} - \alpha_{\rm h})}$, which
is 1.1 keV for the best-fit \chandra\ parameters, and
roughly 0.2 keV and 1 keV for \einstein\ and \rosat,
respectively.
\end{minipage}
\end{table}

\subsection{\chandra}
\label{sec:chandraspectrum}

A fit to a single-component power law finds an absorption that, at more
than 99 per cent confidence, is inconsistently small compared with the
Galactic value.  This is the usual indication of a soft excess.
When the absorption is fixed at the radio-derived value for our Galaxy
we obtain a spectral index of
$\alpha =1.35$ (Table~\ref{tab:xspec}) and an
upper limit to any intrinsic absorption,
in excess of that in our own Galaxy of $5 \times 10^{19}$ cm$^{-2}$.
The fit is only marginally acceptable, at
$\chi^2 = 81$ for 64 degrees of freedom, and shows correlated features
in the residuals which suggest that the model is wrong.
The spectral index is slightly larger than
the value of $0.96\pm 0.04$ tabulated by Siemiginowska et
al. (2003b) for the same observation (no goodness-of-fit
or energy-band information provided),
but the fits are consistent in requiring no intrinsic absorption. 

\begin{figure}
\begin{center}
\epsfxsize 8cm
\epsfbox{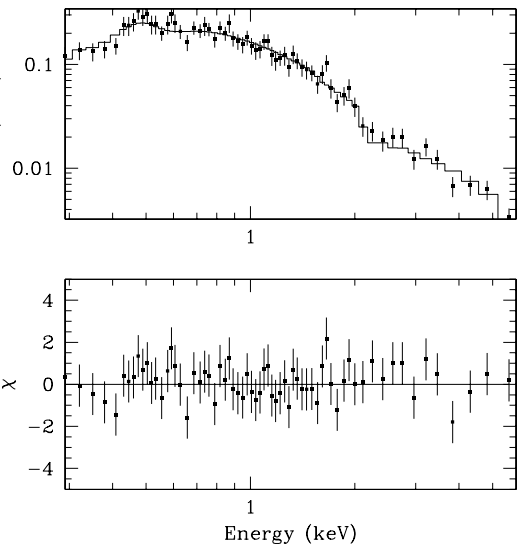}
\end{center}
\caption{\chandra\ spectrum of \css\ extracted from an annulus, as
described in the text.  The data and best-fit two-component power law model
are shown, together with the individual values of $\chi$ contributing to
$\chi^2$.
}
\label{fig:sherpaspec}
\end{figure}

\begin{figure}
\begin{center}
\epsfxsize 8cm
\epsfbox{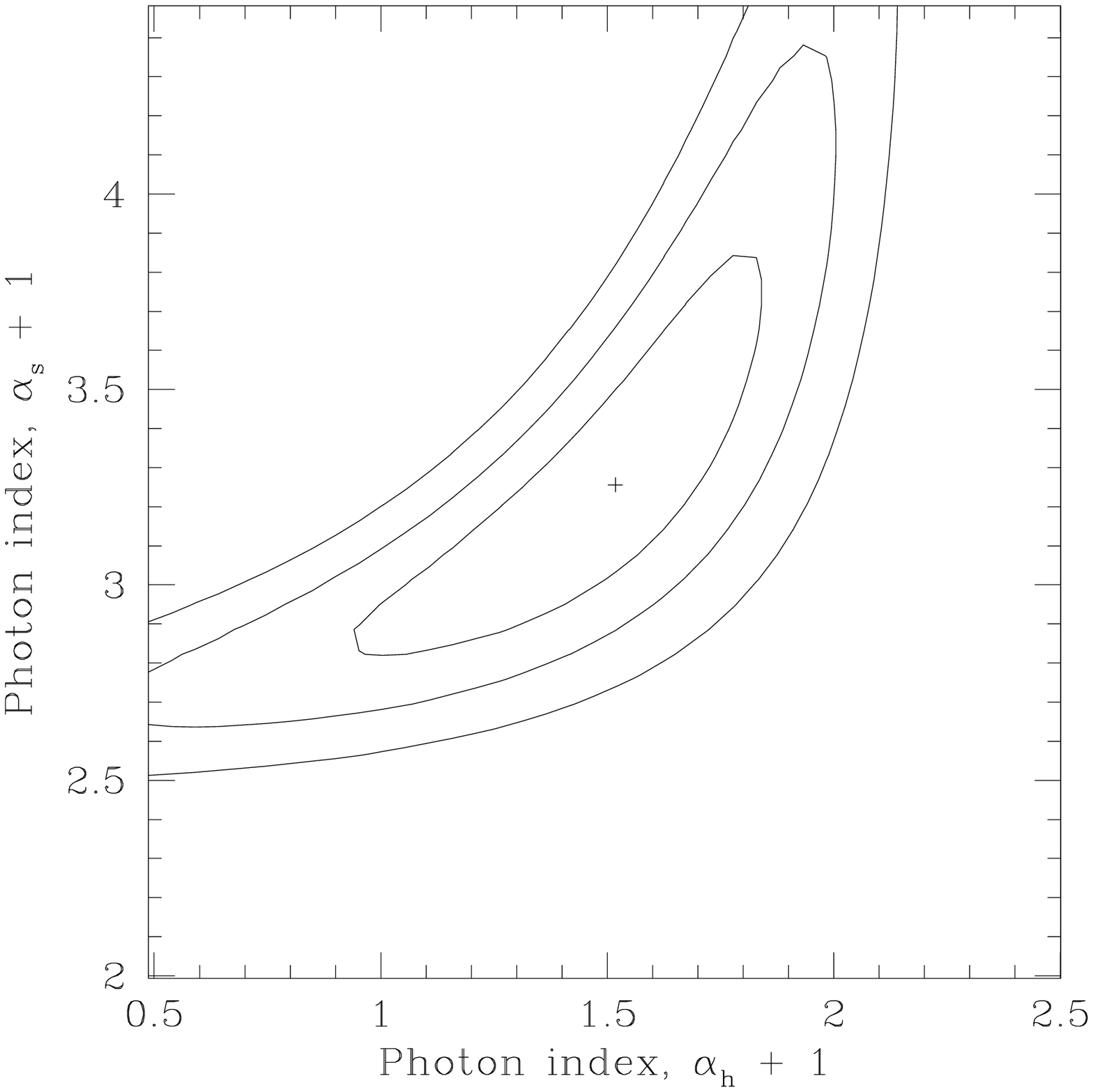}
\end{center}
\caption{Results from fitting a two-component power law to the
\chandra\ data of \css.
The plot shows
contours of equal $\chi^2$ at values of 2.3, 4.6, and 9.2 above the
minimum, corresponding to uncertainties of 68, 90 and 99 per cent
for two interesting parameters.  Note that the plot is of photon
spectral index, one steeper than $\alpha$.
}
\label{fig:sherpacont}
\end{figure}

The addition of a second power law gives $\chi^2 = 42$ for 62 degrees
of freedom, a highly significant improvement on an F test.  The
spectrum, shown in Figure~\ref{fig:sherpaspec}, has best-fit power law
spectral indices of $\alpha_h = 0.5$, $\alpha_s = 2.2$, with
uncertainty contours shown in Figure~\ref{fig:sherpacont}.  The 1 keV
flux densities are $0.34 \pm 0.2$ and $0.4 \pm 0.2$ $\mu$Jy for the
hard and soft components, respectively, where uncertainties are
1$\sigma$ for 2 interesting parameters.  There is no evidence for line
emission from Fe K fluorescence at an observed energy of 4.75 keV in
this or in the spectrum with more counts (but mildly piled up) from
the 2.5 arcsec radius extraction region.

\section{X-ray variability}
\label{sec:xvar}

In \S\ref{sec:rosspectrum} we reported X-ray flux variability between
the January and August PSPC observations (Table~\ref{tab:xobs} and
Figure~\ref{fig:specnorm}). The flux had increased again by the time
of the HRI observation. There is also evidence for variability on a
shorter time scale.  Whilst the August observation was continuous, the
January measurement was made in five intervals.
Table~\ref{tab:xrayvar} lists the start times and durations of the
intervals, together with the 0.2--1.9~keV net counts and count rates
within a circle of radius 2 arcmin centred on \css.  Smaller and
larger on-source regions give similar results.  We find a drop in
count-rate over just 3~hours on January 27 that is comparable to the
average amount of decrease between January and August. Similarly, the
HRI data show variability at the $\sim 20$ per cent level on
time-scales of a few hours.  The flux in the \chandra\ data is at a
similar level to that in the first PSPC exposure.
There is no obvious variability during the \chandra\
observation either below or above 1~keV, 
where the soft and hard components, respectively, dominate the
emission. However, the observation only lasted for 2.6 hours.

\begin{table}
\caption{X-ray variability in the PSPC observations of \css}
\begin{tabular}{lrll}
\hline
Start Time (UTC) & 
Exposure & 
Net Counts & 
Net Count rate \\
& (s) & & (s$^{-1}$)\\
\hline
1992 Jan 26 02:02:45 & 678  & $219\pm16$ &$0.32\pm0.02$\\
1992 Jan 26 03:41:01 & 527  & $193\pm15$ &$0.37\pm0.03$\\
1992 Jan 26 08:24:46 & 748  & $254\pm17$ &$0.34\pm0.02$\\
1992 Jan 27 06:46:45 & 694  & $237\pm17$ &$0.34\pm0.02$\\
1992 Jan 27 09:56:37 & 795  & $198\pm16$ &$0.25\pm0.02$\\
1992 Aug 9 20:03:47 & 2250 & $508\pm24$ &$0.23\pm0.01$\\
\hline
\end{tabular}
\label{tab:xrayvar}
\end{table}

Our results for January 27 correspond to $\Delta L$(0.2--2~keV)
$\approx 1.4 \times 10^{38}$~W in $\Delta t \approx 11.4$~ksec.
$\Delta L/\Delta t$ is above average but not wildly inconsistent with
what has been seen in other quasars and Seyfert galaxies (Barr \&
Mushotzky 1986).  We could interpret the soft excess as the high-energy
tail of unbeamed emission from an accretion disc.
In this case, the
implied source size from the luminosity-doubling
time scale is $2 \times 10^{-4}$ pc ($3 \times 10^{-5}$ mas) which,
for Eddington-limited spherical accretion and an emission region of
size about three times the Schwarzschild radius, implies the presence
of a black hole of about $7 \times 10^8$ M$_\odot$.
In this paper we neither exclude nor explore further the possibility
that all the X-ray emission is associated with the accretion
processes.  Rather we explore possible relations between the radio
and hard X-ray components of the source.
We also discuss briefly in \S \ref{sec:core} the possibility that the X-ray
emission is from an embedded blazar.

\section{\vlba\ results}
\label{sec:vlba}

\begin{figure*}
\epsfxsize 16cm
\epsfbox{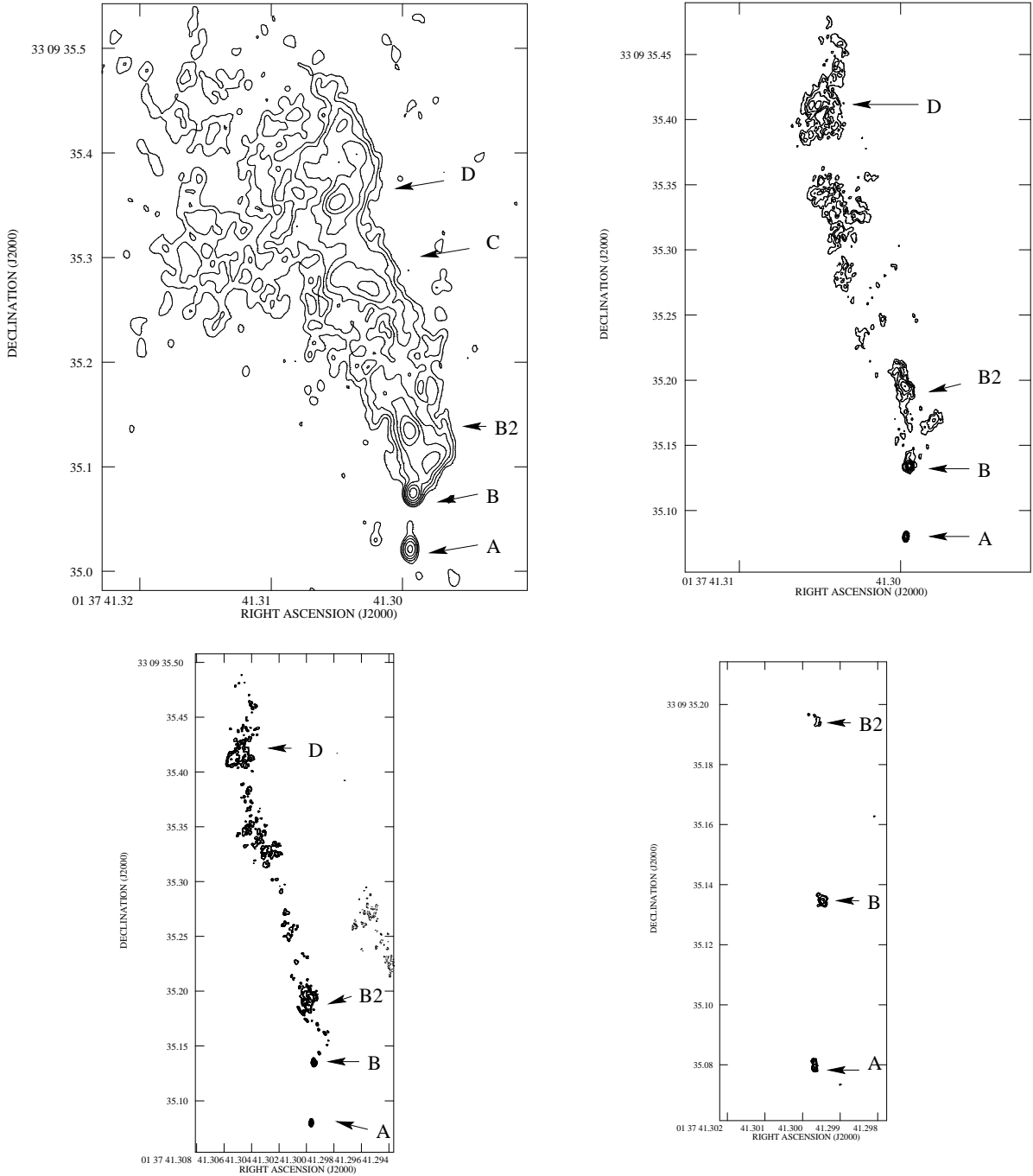}
\caption{\vlba\ maps of \css. Top left: 1.5 GHz, lowest contour 4 mJy
beam$^{-1}$. Top right: 5.0 GHz, lowest contour 2 mJy
beam$^{-1}$. Bottom left: 8.4 GHz, lowest contour 1 mJy
beam$^{-1}$. Bottom right: 15.3 GHz, lowest contour 1 mJy
beam$^{-1}$. Contour levels are logarithmic, increasing by a factor
2. Negative contours are dashed.}
\label{fig:vlbimaps}
\end{figure*}

Final maps at the full resolution available from the data are
presented in Figure~\ref{fig:vlbimaps}. Parameters of the maps are
listed in Table~\ref{tab:parameters}.

\begin{table}
\caption{Parameters of VLBI maps of Figure \ref{fig:vlbimaps}}
\label{tab:parameters}
\begin{tabular}{rrrrrrl}
\hline
Freq.&\multicolumn{3}{c}{Beam}&Off-source&Dyn-&Approx.\\
(GHz)&Major&Minor&Pos-&noise ($\mu$Jy&amic&las\\
     &axis&axis&ition&beam$^{-1}$)&range&(arcsec)\\
&(mas)&(mas)&angle&&&\\
\hline
1.5&10.0&7.47&-1.68&940&250&1.3\\
5.0&3.27&2.28&-1.69&160&670&0.7\\
8.4&2.13&1.45&8.60&122&520&0.4\\
15.4&1.14&1.00&16.07&174&70&0.04\\
\hline
\end{tabular}
\vskip 5pt
\begin{minipage}{8cm}
The approximate largest angular scale (las) describes the largest scale to
which the observations are sensitive, based on the shortest baseline
present in the images.
\end{minipage}
\end{table}

The 1.5-GHz map reproduces well the source structure seen in the
1.6-GHz image of Wilkinson \etal\ (1990). Extending the notation used
in that paper, we will denote the component to the north of `B' `B2'
and the `warm spot' close to the northern edge of the L-band maps
`D'. These components are labelled on each map in
Figure~\ref{fig:vlbimaps}. In a rectangle covering all of the emission
in the high-resolution map of Wilkinson \etal\ (1991), their
map\footnote{Obtained from the radio-source atlas of Leahy, Bridle \&
Strom (http://www.jb.man.ac.uk/atlas/)} contains 9.7 Jy and ours 10.5
Jy.  The small difference between these flux densities is in the sense
and of the magnitude expected given the steep-spectrum nature of
\css.

At higher frequencies much of the structure seen in the 1.5-GHz map is
still visible, although the extended structure is not detectable.
Component C is largely resolved out. B remains compact, but structure
in components B2 and D is clearly present; B2 is resolved into an
elongated structure surrounded by fainter emission while D is resolved
into two components aligned roughly E-W. At 15 GHz the `core'
component A appears to be resolved into two components, possibly a
true core and a jet component, separated by $\sim 3$ mas (20 pc); the
8-GHz core can also be modelled as two Gaussians, but otherwise there
is no high-frequency evidence for the jet connecting A and B suggested
by the maps of Wilkinson \etal , though component A is elongated
northwards in all maps.

The flux densities and deconvolved sizes of the named components are
tabulated in Table \ref{tab:fluxes}. These were obtained with the task
{\sc JMFIT}, fitting a Gaussian together with zero level and slope. To
reduce the effects of the different $uv$ plane sampling, the longest
baselines in the 5, 8.4 and 15-GHz measurements were matched and a
tapered weighting scheme was used to give similar dirty beams.  The
same restoring beam (3.5 mas) was used for all three frequencies so
that in principle the resulting maps should be sensitive to the same
compact structures.  Matching baselines to the 1.5-GHz data was not
feasible; the map of Figure \ref{fig:vlbimaps} was used and we have
tabulated, in parentheses, the flux densities measured when the low
and high observing frequencies are mapped separately. Errors quoted
are those returned by {\sc JMFIT}.

\begin{table}
\caption{Flux densities of named components of Figure \ref{fig:vlbimaps}}
\label{tab:fluxes}
\begin{tabular}{lrrrrr}
\hline
Comp-&Freq.&Flux&Major&Minor&Position\\
onent&(GHz)&density&axis&axis&angle\\
&&(mJy)&(mas)&(mas)&(degrees)\\
\hline
A&1.53&$104\pm 2$&6.9&2.5&179\\
&(1.45&$114\pm 5$&6.1&3.5&9)\\
&(1.61&$100\pm 6$&9.0&2.4&4)\\
&4.99&$56.1\pm 0.3$&2.2&0.4&171\\
&8.42&$53.4\pm 0.2$&2.3&0.3&174\\
&15.4&$21.2\pm 0.4$&2.4&0.2&9\\[8pt]

B&1.53&$243 \pm 1$&5.0&2.8&148\\
&(1.45&$252\pm 5$&4.3&2.6&163)\\
&(1.61&$223\pm 6$&5.5&2.5&136)\\
&4.99&$146.0\pm 0.3$&2.0&1.7&21\\
&8.41&$101.2\pm 0.2$&1.8&1.1&21\\
&15.4&$30.3\pm 0.4$&1.9&0.5&23\\[8pt]

B2&1.53&$547\pm 3$&14.8&8.4&19\\
&(1.45&$571\pm 8$&14.3&8.2&20)\\
&(1.61&$554\pm 10$&15.8&9.2&21)\\
&4.99&$153.0\pm 0.6$&7.3&4.1&43\\
&8.41&$116.6 \pm 0.4$&6.3&4.3&49\\
&15.4&$11.5 \pm 0.5$&5.0&0.7&35\\
\hline
\end{tabular}
\end{table}

Components B and B2 have spectra consistent with a moderately flat
power law ($\alpha \sim 0.5$) at frequencies between 5 and 8.4~GHz,
cutting off at 15 GHz.  For B the power law extrapolates to 1.5 GHz.
For B2 the 1.5-GHz flux density lies above the extrapolation, but this
is not unexpected as the component also appears distinctly larger at
this frequency.  The spectra are consistent with a model in which B
and B2 are physically similar to the hotspots (regions where particles
are accelerated at shocks) of powerful, FRII, radio sources.

For component A, $\alpha$ is roughly 0.6 at 1.5~GHz and between 1.5
and 5 GHz.  Its faintness at 330~MHz (Simon \etal\ 1990) suggest that
the flux density peaks at a frequency between 330~MHz and 1.5~GHz.
The spectrum is very flat between 5 and 8~GHz and falls off rapidly by
15 GHz. A two-component model --- perhaps consisting of a
self-absorbed core and a steep-spectrum jet --- may be necessary to
explain this behaviour.  The core is a sufficiently small contributor
to the overall radio flux of the source that it would need to be
undergoing large fractional variability (of order 100 per cent or more
over time scales of decades) were it to account for the likely
long-term variability in total flux discussed by Ott \etal\ (1994).

A comparison of the flux densities in individual components between
the Wilkinson \etal\ (1991) map and ours is uncertain because the
maximum-entropy algorithm {\sc VTESS} was used in making the former.
Using fixed extraction regions, the components labelled in
Figure~\ref{fig:vlbimaps} all contain a higher flux in the Wilkinson
\etal\ map (ranging from 16 per cent for component A to 53 per cent
for B).  However, due to the different mapping procedures, we are
unable to conclude with any certainty the extent to which the reported
variability in integrated flux (Ott \etal\ 1994) is located in
specific source components.

We used {\sc JMFIT} to search for evidence for proper motion of
components B or B2 with respect to A at 1.5 GHz between the epoch of
Wilkinson \etal 's observations (1984.3) and ours (1996.1). The
usefulness of this measurement is limited by the accuracy with which
we can align the weak components A. The only significant motion is
that of component B, which appears to have moved northwards by $0.3
\pm 0.1$ mas, a $3\sigma$ result. Taken at face value, this
corresponds to an apparent motion of 2 pc in 11.8 years, so that we
have a sub-luminal $v_{\rm app} = (0.5 \pm 0.2)c$.

\begin{figure}
\epsfxsize 6cm
\epsfbox{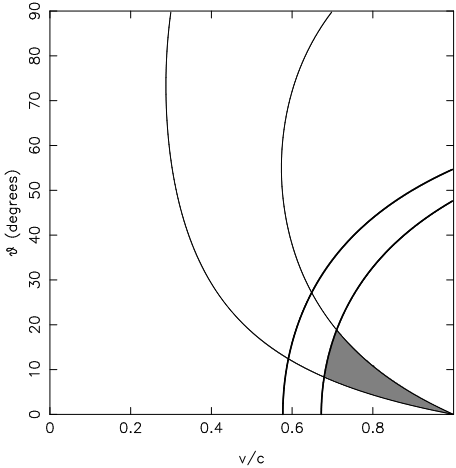}
\caption{Constraints on source orientation and velocity of component
B. The two light lines bracket the region of parameter space allowed
(at the $1\sigma$ level) by the apparent proper motion of B.  The two
heavy lines mark regions allowed by the constraints on source
sidedness: the left-hand line is the locus of points with
jet-counterjet ratio $\sim 100$ (Briggs 1995) while the region to the
right of the right-hand line is allowed by the jet-counterjet ratio
limit determined at knot B. The shaded region is allowed by the
combination of the two constraints.}
\label{fig:const}
\end{figure}

If we assume that the source is intrinsically two sided and symmetric
out to a projected radius of 0.05 arcsec, we can constrain models in
which the sidedness of the jet is related to relativistic beaming.
Based on the off-source noise in the 5-GHz data, we can set a
$3\sigma$ upper limit of $<0.5$ mJy for the flux of a component
corresponding to component B in the hypothetical undetected
counterjet, so that the jet-counterjet flux ratio here is $>300$. This
may be compared to the large-scale jet-counterjet ratio of $\sim 100$
determined from the VLA imaging of Briggs (1995). For an intrinsically
symmetrical pair of emitting regions, the jet-counterjet ratio is
given by

\[
R = \left({{c + {v} \cos \theta}\over{c - v\cos
\theta}}\right)^{3+\alpha}
\]
(the exponent 3 being appropriate for a single emitting region rather
than a jet), while the apparent motion on the sky of the jet component
is given by
\[
v_{\rm app} = {{v c\sin \theta} \over (c-v\cos \theta)}
\]
These results restrict the possible values of $v$ and $\theta$
as shown in Figure~\ref{fig:const}. Only small angles to the line of
sight and relativistic velocities for component B ($v > 0.7c$) are
allowed if the radio structure of \css\ is intrinsically symmetrical
and component B moves with the bulk jet velocity.

\section{The X-ray-emitting atmosphere}
\label{sec:xspat}

An excess of counts (compared with the PSF) in the HRI radial profile of
\css\ on scales of 10--30 arcsec
(70--200 kpc) has been reported
previously (Crawford \etal\ 1999; Hardcastle \& Worrall 1999).
Based on
the improvement in $\chi^2$ when $\beta$  models of fixed parameters
were added to a point-source model, it was inferred that \css\ lies in a luminous
cluster of galaxies.  When we adopted the method applied by Worrall \&
Birkinshaw (2001) to 3C\,346, in order to determine the uncertainty in
the derived parameters, we found allowed values of $\beta =
0.87^{+0.73}_{-0.19}$, $\theta_c = 8.7^{+10.7}_{-4.5}$ (1$\sigma$
errors), with the unresolved core contributing $5646^{+136}_{-484}$ of the
$6000\pm 166$ total counts within a source-centred circle of radius of 2.5
arcmin.

There is a recent concern about the accuracy of the X-ray luminosities
of clusters around 3CR quasars measured with the HRI
(Crawford \etal\ 1999; Hardcastle \& Worrall 1999).  
Crawford \& Fabian (2003) report \chandra\ observations for
three 3CR quasars with earlier HRI cluster detections, and in two of
them the \chandra-measured luminosities are significantly lower than
the HRI-measured values (by a factor of 20 to 50).  Crawford \& Fabian
attribute this discrepancy to insufficient or erroneous wobble
corrections affecting the HRI results. 
That this may be
an issue for \css\ is suggested by the relatively low (but
uncertain) quasar-galaxy spatial covariance found by Yee \& Green
(1987).  

Indeed we find that \chandra\ does not support the existence of
a cluster as luminous as found in the HRI data.
Using only the best-fit HRI-derived values of
$\beta$ and $\theta_c$, the fit of the \chandra\ radial profile to
a two-component model gives $\chi^2$=8.9 for 21 degrees of freedom,
as compared with  $\chi^2$=10.3 for 22 degrees of freedom for a
point-like component alone, such that the probability of the
improvement being by chance is 8.5 per cent on an F test.
No other choice of $\beta$ and $\theta_c$ gives a detection that can
be claimed with greater confidence, where searching through these
parameters further decreases the number of degrees of freedom by two.
So, the detection of any extended emission is marginal.

If we assume that the \chandra\ detection of extended emission is
real, the component contains only $30\pm 6$ counts and is a factor of
40 weaker than predicted by the HRI.  If we further assume that the
emission is from a hot gaseous environment of the quasar, the bolometric
X-ray luminosity is $\sim 2 \times 10^{36}$ W, assuming a temperature
of $kT = 1.5$ keV so that it lies on the cluster
temperature-luminosity correlation (Arnaud \& Evrard 1999).  We
convert the X-ray parameters to estimates of physical conditions using
the approach of Birkinshaw \& Worrall (1993).  The mass deposition
rate from cooling would fall far short of that predicted by Fabian
\etal\ (1987)'s explanation for the emission-line nebulosity.  The gas
pressure at a radius of 6 arcsec ($\sim 40$ kpc) is $\sim 2 \times
10^{-12}$~Pa, which is insufficient to confine the extended
emission-line regions if their pressures are $\sim 10^{-11}$ Pa, as
estimated by Fabian \etal\ (1987) on the basis of a photoionization
model.

At a radius of about 0.5 arcsec, within which
most of the radio structure is contained, the cluster pressure would
be about $3 \times 10^{-12}$~Pa.  The plume of 1.5-GHz radio emission
stretching north from component B for about 0.3 arcsec, and contained
within continuous contours in Figure~\ref{fig:vlbimaps}, has a flux
density of 8.7~Jy (roughly half of the total emission from \css\ at
this frequency).  Using an electron spectrum of the shape and energy
limits described in \S \ref{sec:radio-related} and
Table~\ref{tab:icftab}, the minimum pressure in radiating particles
and magnetic field (for no proton contribution) is found to be roughly
$2 \times 10^{-9}$ Pa if the source is unbeamed.  Thus despite the
diffuse appearance, which does not suggest a strongly overpressured,
rapidly expanding source, the external X-ray-emitting gas falls well
short of providing pressure balance.  The minimum internal pressure
decreases if beaming is important.  For models consistent with
Figure~\ref{fig:const}, the minimum conditions for pressure balance
are only met if the plume velocity is greater than $0.985$c and it is
at angle to the line of sight of less than 0.4 degrees.  X-ray
constraints from \rosat\ observations of two GPS galaxies have already
found pressures one to two orders of magnitude too low to confine
their radio sources assuming there is no beaming (O'Dea \etal\ 1996),
and in a sample of FRIIs, Hardcastle \& Worrall (2000) find similar
results for the lobes of the five sources of maximum linear size $\leq
10$~kpc

\section{Radio-related X-ray emission}
\label{sec:radio-related}

\begin{figure}
\epsfxsize 7cm
\epsfbox{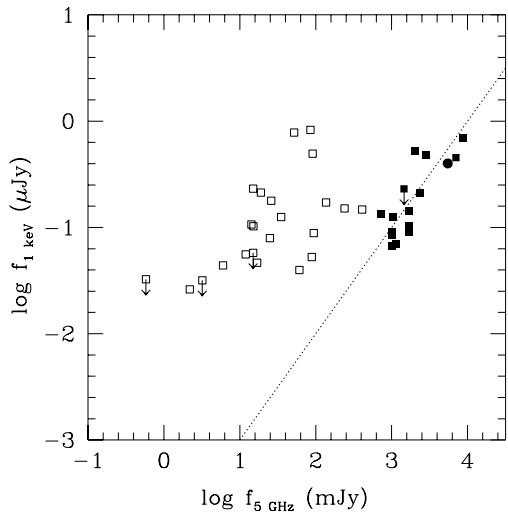}
\caption{Core X-ray and radio flux densities for quasars with a spread
of about 2 orders of magnitude in isotropic radio power and with $z >
0.3$, from Worrall \etal\ (1994).  There is a correlation of slope
unity for the core-dominated quasars (filled squares) but not the
lobe-dominated sources (open squares).  \css, shown as a filled
circle, is consistent with the correlation for core-dominated sources
when its hard X-ray emission of $0.4\mu Jy$ and all its radio emission
(on core and jet scales) of flux density 5.48 Jy (VLA
calibrator manual: http://www.aoc.nrao.edu/$\sim$gtaylor/csource.html)
is included.} 
\label{fig:xrplot}
\end{figure}

The radio, X-ray, and optical luminosities of samples of quasars have
been investigated using a variety of different approaches, leading
most authors to conclude that emission associated with the compact
radio source dominates the X-radiation in core-dominated radio
sources, while a ubiquitous unbeamed component takes over as the
dominant source of X-ray emission in lobe-dominated sources (e.g.,
Zamorani 1984; Browne \& Murphy, 1987; Worrall \etal\ 1987; Kembhavi
1993).  This can explain why core-dominated radio-loud quasars exhibit
an approximately linear correlation between their 1-keV X-ray and
5-GHz radio flux densities, while lobe-dominated quasars tend to lie
above the line (Fig.~\ref{fig:xrplot}).

\css\ is interesting in this context. Using the hard
non-variable component of X-ray emission and the total radio emission
(which is all compact on the scale of the other quasars)
it lies close to the
line defined by other core-dominated quasars
(Fig.~\ref{fig:xrplot}). If the proximity of \css\ to the correlation
is not coincidental, then we expect a substantial fraction of the
hard X-ray emission to be produced by the radio-emitting
components.  In the remainder of this paper we test this possibility
by estimating the X-ray emission from non-thermal processes in
the radio components that we resolve.  We also explore briefly
in \S\ref{sec:core}
the possibility that the X-ray emission arises from an embedded
blazar that contributes negligibly to the observed radio emission.

Radio components can give rise to X-ray emission by two main
mechanisms; synchrotron and inverse-Compton (IC) emission. There is
clear evidence for the importance of both of these processes in extended
components of radio sources studied with \chandra. In particular, the
jets of low-luminosity, FRI, radio galaxies (which in many respects
resemble the \css\ jet) are often synchrotron emitters in the X-ray
(e.g., Worrall, Birkinshaw \& Hardcastle 2001; Hardcastle, Birkinshaw \&
Worrall 2001) while X-ray emission likely to be due to the
inverse-Compton process has been detected from the jets, hotspots and
lobes of a number of powerful FRII sources (e.g., Sambruna \etal\
2002; Hardcastle \etal\ 2002).

If the X-ray emission is of an IC origin, the spectral index should
match that of the optically-thin radio emission, which is true for the
hard X-ray component (Fig.~\ref{fig:sherpacont}) in both
the plume, where $\alpha \approx 0.75$ (Ott \etal\ 1994),
and the knots (Table~\ref{tab:fluxes}).
The variable soft X-ray component would require a separate
explanation, such as the high-energy tail of the emission from an
accretion disc (\S\ref{sec:xvar}).

Estimates of the IC emission necessarily rely on assumptions about the
geometry of the radio components, their bulk speed, and the magnetic
field strength in the radio-emitting plasma. With a sufficiently low
magnetic field strength, any radio-emitting component can give the
required IC X-ray flux. However, there is now evidence that the
extended components of radio sources have magnetic field strengths
consistent with, or a factor of a few lower than, the values for
energy equipartition between the radiating electrons and the magnetic
field (e.g., Brunetti \etal\ 1999; Hardcastle \etal\ 2002). A magnetic
field which is a factor of a few below the equipartition value can
increase by an order of magnitude the predicted X-ray emissivity.

For IC calculations we must also make some assumptions about the
electron spectrum.  We model this as a single power law, $N(\gamma)
\propto \gamma^{-p}$ between an upper and lower cutoff energy,
parametrized in terms of the lower and upper limits on the Lorentz
factors, $\gamma_{\rm min}$ and $\gamma_{\rm max}$. We take $p$ to be
$2.5$, reproducing $\alpha \sim 0.75$ observed in the total radio
emission. The normalization and values of $\gamma_{\rm min}$ and
$\gamma_{\rm max}$ are selected based on available constraints from
the broad-band synchrotron spectrum in the region of interest.  We use
the anisotropic IC code described by Hardcastle \etal\ (2002) for our
calculations.  This deals accurately with the effects of beaming and
aberration on the illuminating photon population and the IC emission
itself.  The three parent photon populations we will consider are (i)
the synchrotron photons themselves (SSC), (ii) the cosmic microwave
background radiation (CMB/IC), and (iii) radiation from the active
nucleus (NIC).

The NIC model requires some additional assumptions.  For no beaming
and an equipartition magnetic field, the electron
population responsible for the observed radio emission produces X-ray
IC most successfully through upscattering far infra-red photons.  We
use the \iso\ results of Meisenheimer \etal\ (2001), and model the
spectrum as having a flux density of 829~mJy at $3 \times 10^{12}$~Hz
(observed frame), above which frequency $\alpha \approx 1.0$.  The
results are insensitive to the upper-frequency cut-off but are
sensitive to the photon spectrum below $3 \times 10^{12}$~Hz.  The
Meisenheimer \etal\ data show that the spectrum flattens, and are
consistent with $\alpha$ between $\sim 0.0$ and $-1.0$ down to a
frequency of $1.7 \times 10^{12}$~Hz.  Photons of lower frequency
still, in the range between $1.7 \times 10^{12}$~Hz and $3 \times
10^{11}$~Hz, could in principle boost the IC emission from this
process considerably, but there is no justification for the
introduction of a new soft population of photons in this unobserved
wavelength band.  We cut off the spectrum at $1.7 \times 10^{12}$~Hz
but adopt $\alpha=0$ below $3 \times 10^{12}$~Hz.  The photons are
allowed to illuminate the radio-emitting regions as though they
originated at a point source at the quasar nucleus. Because some of
the IR emission in \css\ may originate in a circumnuclear starburst,
this model is not necessarily accurate.  However, as long as the
starburst emission does not come from a region much larger than the
radio source, it is not seriously in error.

We consider separately the resolved radio emission and the radio core
as possible sites of radio-related X-ray emission.  The former may
explain the position of \css\ in Figure~\ref{fig:xrplot}.

\subsection{Jet-related X-rays}
\label{sec:jets}

The main difficulty with a synchrotron model in which both radio and
X-ray emission are from a population of electrons with a
single-component spectrum (power law or broken power law) is the upper
limit on jet surface brightness of 22.2 mag arcsec$^{-2}$ in the \hst\
F814W filter found by Kirhakos \etal\ (1999).  This implies an upper
limit of order 1 $\mu$Jy to the integrated jet emission at a frequency
of $\sim 3.8 \times 10^{14}$ Hz.  Since this upper limit is comparable
to the measured X-ray flux density at 1~keV, synchrotron emission
could only make a significant contribution to the unresolved X-rays
through a highly non-standard electron population which produces
copious emission at frequencies only above the optical.  The detection
of optical jet emission is severely hampered by the bright nucleus and
galaxy nebulosity, but in view of the Kirhakos \etal\ constraint we
consider the possibility of significant X-ray synchrotron emission
from the radio jet no further.

The brightest small-scale components in the jet are B and B2.
However, these components contain less than 10 per cent of the radio
flux of the overall jet, and, assuming source regions of about 2~mas
in extent, none of the photon populations gives more IC X-ray emission
than we find from the overall radio plume.

We model the plume described in \S\ref{sec:xspat} crudely as a
truncated cone, with a linear gradient in electron density along the
cone axis to represent the brightness gradient along the plume seen in
the 1.5-GHz maps. The total 1.5-GHz flux density of 8.7~Jy for this
volume provides the normalization for the electron spectrum.  A single
constant magnetic field strength is assumed throughout the source for
numerical efficiency.  We assume that the spectral shape of the plume
matches that of the integrated radio emission and by default use the
$\sim 80$ MHz turnover frequency (plausibly due to synchrotron self-absorption or
free-free absorption; Wilkinson \etal\ 1990) to provide a limit
on the low-energy cutoff of the electrons. The spectrum also steepens
before mm-wave radiation is emitted (Steppe \etal\ 1995), which gives
an indication of a possible high-energy cutoff. The corresponding
values of $\gamma_{\rm min}$ and $\gamma_{\rm max}$
(Table~\ref{tab:icftab}) depend on the degree to which the extended
emission is beamed and on the derived equipartition magnetic field strength.

\subsubsection{Unbeamed plume model}
\label{sec:ubplume}

If the plume is unbeamed and lies in the plane of the sky, the
one-sidedness of the jet is a genuine intrinsic feature of the source,
perhaps caused by differences in the beam's interaction with the
external medium, as suggested by Wilkinson \etal\ (1990).  The
predicted SSC, CMB/IC and NIC X-ray flux densities, assuming
equipartition, are given in Table~\ref{tab:icftab}. The combined
output is a factor of $\sim 50$ below the observed flux density of
$\sim 340$ nJy (Table~\ref{tab:xspec}) in the
hard X-ray component.  We would require the magnetic field to be a
factor of about 10 below the equipartition value if the mechanisms are
to make an important contribution.  A reduction to only a factor of
six below equipartition is sufficient if $\gamma_{\rm min}$ is reduced
to 5, allowing for the presence of electrons at energies below that
for which the synchrotron emission is self-absorbed or suffers
free-free absorption.  The favoured interpretation of the resolved
X-ray jet emission from the quasar jet PKS 0637-752 supports the
presence of electrons with $5 < \gamma_{\rm min}< 20$ (Tavecchio
\etal\ 2000).

\begin{table}
\caption{Predicted X-ray emission from IC models applied to the plume,
assuming equipartition}
\label{tab:icftab}
\begin{tabular}{lrrlllr}
\hline
Model&$\gamma_{\rm min}$&$\gamma_{\rm max}$&\multicolumn{4}{c}{IC flux
density at 1 keV (nJy)}\\ 
&&&SSC&CMB/IC&NIC&Total\\
\hline
Unbeamed&180&7100&3.3&0.003&3.5&7\\
Beamed&200&8000&0.04&0.9&0.0002&1\\
\hline
\end{tabular}
\end{table}

The NIC contribution would be increased if the IR radiation were
increased through beaming.  To contribute the observed X-rays, assuming
equipartition,  a component
of beamed IR emission would need both to be about a factor of 100
brighter than the measured IR (as seen by the jet) and have a rather
large opening angle (Figure~\ref{fig:vlbimaps}) to cover the whole jet. 
Although such extra IR emission
cannot be ruled out, there are no observations of which we are aware
which support the presence of dominant beamed IR emission.  On the
contrary, Meissenheimer \etal\~(2001) argue that radio galaxies and quasars
have similar IR luminosities, implying that the emission seen is
emitted isotropically.  To get a maximum gain from scattering
geometry, the nuclear beamed
emission should be directed away from the line of sight.  It then
seems most likely that equal and opposite beamed emission should
dominate the IR which is observed, and this is insufficient to produce
the observed X-rays.  We thus consider it unlikely that beamed
IR radiation is the dominant photon source for producing the X-rays.

\subsubsection{Beamed plume model}
\label{sec:bplume}

In Figure~\ref{fig:const} we presented the range of beaming parameters
that are consistent with the observed sidedness and the possible
detection of sub-luminal motion. As a representative example we take
$v = 0.9c$ and $\theta = 5$ degrees, giving a bulk relativistic
Doppler factor of $\delta = 4.2$. We take the projection of the
velocity vector on the sky to point along the direction of apparent
motion of component B, i.e. approximately N on the radio map.  In this
model, the boosted CMB radiation is the dominant photon population in
the source frame (Table~\ref{tab:icftab}). Inverse-square law and
Doppler effects greatly reduce the effects of the nuclear
illumination. The resulting total predicted X-ray emission is two
orders of magnitude below the observed value. The magnetic field would
have to be considerably below the equipartition value to achieve
results approaching the observed X-ray flux, and we conclude that in
this beamed model the plume is highly unlikely to contribute to any
significant extent to the observed X-rays.

We can ask how large the beaming would need to be to increase the
CMB/IC prediction to the observed level.  The result is that we would
require all the electrons in the plume to have $v = 0.994c$, $\theta =
0.17$ degrees, $\delta = 18$.  
Unless the source has an unusually small intrinsic opening angle which
is maintained for hundreds of kpc close to the line of sight,
only a fraction of the electrons will participate in such boosting,
and these parameters represent a lower
limit to the required beaming.  We thus conclude that it is unlikely
that all the observed hard X-ray emission is from this mechanism.

\subsection{Core-related X-rays}
\label{sec:core}

A possible site for radio-related X-ray emission (although not one
which alone would explain the position of \css\ in
Figure~\ref{fig:xrplot}) is emission from within component A.  Here
the NIC can be high if the radio component is unbeamed.  For example,
modelling the photon and electron spectra as above, and providing the
photon illumination from a point source embedded in the base of a
cylindrical jet of radius 0.25~mas (from Gaussian fits to two
components) and length 4~mas (from measurement), we predict X-ray
emission of 20~nJy at 1~keV, somewhat above that from the extended
plume. Our modelling of the far-infrared radiation source as
point-like to component A is probably consistent with Meisenheimer
\etal\ (2001)'s interpretation as originating from the type of dusty
torus required by models unifying radio galaxies and quasars.
However, if the source of this radiation is larger than component A,
or if it is point-like and the quasar nucleus is south of component A,
the predicted X-ray flux will be lower.  Taking the adopted geometry
at face value, it is possible to make the hard X-ray component with a
magnetic field strength only a factor of two to three below
equipartition.  This is in line with the level of field deduced from
X-ray inverse Compton measurements in components of powerful radio
galaxies and quasars on larger scales (e.g., Brunetti et al. 1999;
Hardcastle \etal\ 2002).  Our modelling finds a declining X-ray to
radio ratio with distance from the nucleus when knot A and the plume
are compared under similar equipartition assumptions.

\begin{figure}
\begin{center}
\epsfxsize 8cm
\epsfbox{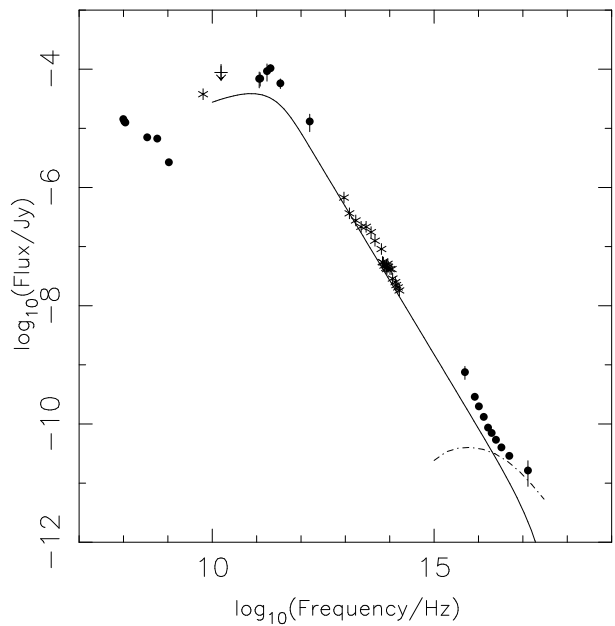}
\end{center}
\caption{Data for the core of \css: radio (component A) 
and soft X-ray component
from this work, far-IR from Meisenheimer \etal\ (2001), other data from
Elvis \etal\ (1994). No reddening
corrections are applied to the data. Data are shifted to the source
rest frame.  The model is synchrotron (solid line) and self-Compton
(dot-dashed) assuming a source of size $10^{16}$ cm, beamed by $\delta
= 20$ ($\beta=0.995$, $\theta =0.144$ degs).  The minimum electron
Lorentz factor is 200 since the source would be self-absorbed below
this.  The electron spectral slope is 3.5, i.e., a
steepening of 1.5 from a canonical slope, but consistent with that
measured above the break in lower-power radio sources like M~87
(B\"ohringer \etal\ 2001). The SSC calculation
assumes equipartition between the electrons and magnetic field, and
the equipartition magnetic field is $1.3 \times 10^{-4}$ T (1.3
Gauss).  The synchrotron energy loss timescale for the lowest-energy
electrons is 27 days (rest frame) and scales as the inverse of the energy.
}
\label{fig:blazar}
\end{figure}

While the fast variability seen at soft X-ray energies with
\rosat\ may suggest unbeamed emission from the high-energy spectral
tail of an accretion disc, we can consider the alternative possibility that we
are seeing a somewhat larger beamed component embedded in the core.
Simple models of a spherical moving blob have proved relatively
successful in describing the multiwavelength spectra of blazars up to
gamma-ray (and in some cases TeV) energies via SSC or inverse Compton
scattering of external photons (e.g., Ghisellini et al. 1998).  The
energy at which the Compton-scattered emission becomes dominant as
a hard component is normally in the X-ray
band, and so may explain the concave spectrum that we observe.
In Figure \ref{fig:blazar} we plot the spectrum of the core,
together with an example SSC model.  We have chosen beaming
parameters consistent with the constraints at larger radii (assuming a
symmetrical jet) of Figure \ref{fig:const}, and a source size which is
consistent with the observed soft X-ray variability.  The model
parameters are selected to describe roughly the level of X-ray
emission while not overproducing emission at lower energies. The radio
emission, much of the far-IR emission, and the blue bump are not
described by the model, and the position of \css\ in Figure
\ref{fig:xrplot} is not explained.  Measurements of gamma rays from
\css\ could help establish the credibility of an embedded blazar
explanation for the X-ray emission.

\section{Conclusions}

The X-ray spectrum of \css\ is the composite of a soft, steep-spectrum
component and harder, power-law emission of slope consistent with the
1~GHz radio spectrum.
The variability measured with \rosat\ suggests that at least part of
the X-ray emission (principally the softest) arises from a region of
accretion-disc size or from a beamed blazar component.

The \chandra\ data give only a marginal detection of extended
emission.  Even if this component is real and due to cluster gas,
the radio plume is over-pressured unless it is beamed towards us at an
angle of less than 0.4 degrees to the line of sight and a bulk velocity
of more than 0.985c.  The gas pressure at the radius of the extended
emission-line regions is at most $\sim 2 \times 10^{-12}$ Pa.

We have used our new multifrequency \vlba\ data to study the
relationship between the X-ray and radio components.  There is no
excess absorption in the X-ray spectrum which might have allowed us to
conclude that \css\ is small (most of the emission is within 0.5
arcsec of the nucleus) because it is bottled up by cold neutral gas.

We find weak ($3\sigma$) evidence for a proper motion of
$(0.5\pm0.2)c$ in the compact radio component B, about 0.05 arcsec
from the core.  It is quite plausible that the radio emission from
\css\ undergoes sufficient interaction with the external medium that
the source is intrinsically one sided at such distances from the
nucleus.  In that case relativistic beaming may not be important.
However, if we assume intrinsic symmetry, the upper limit to
the flux of knot B's counterjet and the measured proper motion combine
to suggest that relativistic boosting along the line of sight is
important.

The X-ray emission of core-dominated radio-loud quasars is believed to
be dominated by non-thermal radiation from the radio core, and indeed
these sources show a good correlation between radio and X-ray flux.
The fact that \css\ is consistent with the correlation leads to an
expectation that substantial X-ray emission should arise from its
radio features.  At VLBI resolution, \css\ is unusual as compared with
flat-spectrum quasars in that less than 1 per cent of its 1.5-GHz flux
density arises from the core, and so X-ray emission from these
extended radio regions would be required.  The radio source is so
compact that even \chandra's resolution is insufficient to resolve
spatially any X-ray emission from the radio structures.  However,
using the X-ray spectral results we find that synchrotron-emitting
electrons in the radio plume could contribute significantly to the
harder X-ray emission through SSC and up-scattering infra-red emission
from the quasar nucleus.  This requires the radio-regions to be free
from relativistic boosting, in which case we would infer that the
radio source is intrinsically one sided at distances of 300 pc and
more from the nucleus, and it requires that the magnetic field is a
factor of six to ten below the equipartition value.  Stringent
requirements on the jet opening angle make it unlikely that all of the
X-ray emission is from a fast jet which sees boosted CMB emission and
emits beamed X-rays in the observer's frame, the model which is
currently the most popular for explaining resolved X-ray emission from
large-scale quasar jets.

The possibility remains that the X-ray emission from \css\ arises from
close to the central engine.  It may be unrelated to the radio
emission, either from an embedded blazar or associated with the
accretion processes, or the hard X-ray emission may arise from the
up-scattering of infra-red emission from the quasar nucleus if the
geometry is favourable and the magnetic field in the inner region is a
factor of two to three below equipartition.  The fact that the ratio
of X-ray to radio emission is so close to that in quasars where a
radio-X-ray relationship is established is then a coincidence.

\section*{Acknowledgements}

We acknowledge support from NASA contract NAS8-39073 and grant NAG
5-1934 during the early stages of this work, and DMW thanks the
Caltech Astronomy Department for hospitality.  MJH thanks the Royal
Society for a Research Fellowship. We are grateful to the
anonymous referee for suggesting we consider the emission from a
beamed component within the core and for other suggestions which improved
the paper, and to Mark Birkinshaw for discussions.  The VLBA is an
instrument of the National Radio Astronomy Observatory, a facility of
the National Science Foundation operated under cooperative agreement
by Associated Universities, Inc.


\begin{thebibliography}{}
\bibitem{ae} Arnaud, M., Evrard, A.E., 1999, MNRAS, 305, 631
\bibitem{axon} Axon, D.J., Capetti, A., Fanti, R., Morganti, R.,
Robinson, A., Spencer, R., 2000, AJ., 120, 2284
\bibitem{barr} Barr, P., Mushotzky, R.F. 1986, Nature, 320, 421
\bibitem{barthel} Barthel, P.D., Pearson, T.J., Readhead, A.C.S.,
1988, ApJ, 329, L51
\bibitem{birk} Birkinshaw, M., 1994, in Crabtree, D.R., Hanisch, R.J.,
Barnes, J., eds., ASP Conference Series, Vol.~61, p.~249
\bibitem{birk2} Birkinshaw, M., Worrall, D.M., 1993, ApJ, 412, 568
\bibitem{boeh} B\"ohringer, H., et al., 2001, A\&A 365, L181
\bibitem{briel} Briel, U.G. \etal\ 1994, The ROSAT
Users'  Handbook, Max-Planck-Institute f\"ur Extraterrestrische
Physik
\bibitem{briggs} Briggs, D.E., 1995, Ph.D.~thesis, New Mexico
Institute of Mining and Technology
\bibitem{browne} Browne, I.W.A., Murphy, D.W., 1987, MNRAS, 226, 601
\bibitem{brunetti} Brunetti, G., Comastri, A., Dallacasa, D., Bondi,
M., Pedani, M., Setti, G., 1999, A\&A, 342, 57
\bibitem{chatz} Chatzichristou, E.T., Vanderriest, C., Jaffe, W.,
1999, A\&A, 343, 407
\bibitem{cotton} Cotton, W.D., Dallacasa, D., Fanti, C., Fanti, R.,
Foley, A.R., Schilizzi, R.T., Spencer, R.E., 1997, A\&A, 325, 493
\bibitem{craw} Crawford, C.S., Lehmann, I., Fabian, A.C., Bremer,
M.N., Hasinger, G., 1999, MNRAS, 308, 1159
\bibitem{craw2} Crawford, C.S., Fabian, A.C., 2003, MNRAS, 339, 1163
\bibitem{davis} Davis, J.E., 2001, ApJ, 562, 575
\bibitem{elvis} Elvis, M., Lockman, F.J., Wilkes, B.J., 1989, AJ, 97, 777
\bibitem{elvis2} Elvis, M. et al., 1994, ApJS, 95, 1
\bibitem{fabian} Fabian, A.C., Crawford, C.S., Johnstone, R.M., 
Thomas, P.A., 1987, MNRAS, 228, 963
\bibitem{fanti} Fanti, R., Fanti, C., Schilizzi, R.T., Spencer, R.E., 
Nan, R., Parma, P., van Breugel, W.J.M., Venturi, T.,
1990, A\&A, 231, 333
\bibitem{gamb} Gambill, J.K., Sambruna, R.M., Chartas, G., Cheung,
C.C., Maraschi, L., Tavecchio, F., Urry, C.M., Pesce, J.E., 2003,
A\&A, 401, 505
\bibitem{ghis} Ghisellini, G., Celotti, A., Fossati, G., Maraschi, L.,
Comastri, A., 1998, MNRAS, 301, 451
\bibitem{green} Greenstein, J.L., Matthews, T.A., 1963, Nature, 197, 1041
\bibitem{hard} Hardcastle, M.J., Worrall, D.M., 1999, MNRAS, 309, 969
\bibitem{hard2} Hardcastle, M.J., Worrall, D.M., 2000, MNRAS, 319, 562
\bibitem{hbw} Hardcastle, M.J., Birkinshaw, M., Worrall, D.M., 2001,
MNRAS, 326, 1499
\bibitem{hlobe} Hardcastle, M.J., Birkinshaw, M., Cameron, R.A.,
Harris, D.E., Looney, L.W., Worrall, D.M., 2002, ApJ, 581, 948
\bibitem{harris98} Harris, D.E., Silverman, J.D., Hasinger, G.,
Lehman, I., 1998, A\&AS, 133, 431
\bibitem{hartas} Hartas, J.S., Rees, W.G., Scott, P.F., Duffett-Smith,
P.J., 1983, MNRAS, 205, 625
\bibitem{kem} Kembhavi, A., 1993, MNRAS, 264, 683
\bibitem{kirh} Kirhakos, S., Bahcall, J.N., Schneider, D.P., Kristian,
J., 1999, ApJ, 520, 67
\bibitem{kus90} Kus, A.J., Wilkinson, P.N., Pearson, T.J., Readhead,
A.C.S., 1990, in Zensus J.A.~\& Pearson, T.J., eds., Parsec-Scale
Radio Jets, Cambridge University Press, p.~161
\bibitem{kus93} Kus, A.J., Booth, R.S., Marecki, A., Maszkowski, R.,
Porcas, R.W., Pearson, T.J., Readhead, A.C.S., Wilkinson, P.N., 1993, 
in Davis, R.J.~\& Booth, R.S., eds., Sub-arcsec Radio Astronomy
Cambridge University Press, p.~222
\bibitem{LRL}Laing, R.A., Riley, J.M., Longair, M.S., 1983, MNRAS, 204, 151
\bibitem{matt} Matthews, T.A., \& Sandage, A.R., 1963, ApJ, 138, 30
\bibitem{meis} Meisenheimer, K., Haas, M., M\"uller, S.A.H., Chini,
R., Klaas, U., Lemke, D., 2001, A\&A, 372, 719
\bibitem{nan} Nan, R., Schilizzi, R.T., Fanti, C., Fanti, R.,
1991, A\&A, 252, 513
\bibitem{neug} Neugebauer, G., Soifer, B.T., Miley, G.K., 1985, 
ApJ, 295, L27
\bibitem{odea} O'Dea, C.P., 1998, PASP, 110, 493
\bibitem{odea2} O'Dea, C.P., Worrall, D.M., Baum, S.A., Stanghellini,
C., 1996, AJ., 111, 92
\bibitem{ott} Ott, M., Witzel, A., Quirrenbach, A., Krichbaum, T.P., 
Standke, K.J., Schalinski, C.J., Hummel, C.A., 1994, A\&A, 284, 331
\bibitem{pw82} Peacock, J.A., Wall, J.V., 1982, MNRAS, 198, 843
\bibitem{prieto} Prieto, M.A., 1996, MNRAS, 282, 421
\bibitem{samb} Sambruna, R.M., Maraschi, L., Tavecchio, F., Urry,
C.M., Cheung, C.C., Chartas, G., Scarpa, R., Gambill, J.K., 2002, ApJ,
571, 206
\bibitem{scov} Scoville, N.Z., Padin, S., Sanders, D.B., 
Soifer, B.T., Yun, M.S., 1993, ApJ, 415, L75
\bibitem{shast} Shastri, P., Wilkes, B.J., Elvis, M., McDowell, J., 
1993, ApJ, 410, 29
\bibitem{siema} Siemiginowska, A., Aldcroft, T.L., Bechtold, J.,
Elvis, M., 2003a, ApJ, in preparation
\bibitem{siemb} Siemiginowska, A., Aldcroft, T.L., Bechtold, J.,
Brunetti, G., Elvis, M., Stanghellini, C., 2003b, PASA, 20, 113
\bibitem{sim83} Simon, R.S., Readhead, A.C.S., Moffet, A.T.,
Wilkinson, P.N., Allen, B., Burke, B.F., 1983, Nature, 302, 487
\bibitem{simon} Simon, R.S., Readhead, A.C.S., Moffet, A.T., Wilkinson,
P.N., Booth, R., Allen, B, Burke, B.F., 1990, ApJ, 354, 140
\bibitem{steppe} Steppe, H., Jeyakumar, S., Saikia, D.J., Salter,
C.J., 1995, A\&AS, 113, 409
\bibitem{stock} Stockton, A., Ridgway, S.E., 1991, AJ, 102, 488
\bibitem{tav} Tavecchio, F., Maraschi, L., Sambruna, R.M., Urry, C.M.,
2000, ApJ, 544, L23
\bibitem{wilkes} Wilkes, B.J., Elvis, M., 1987, ApJ, 323, 243
\bibitem{wilk90} Wilkinson, P.N., Tzioumis, A.K., Akujor, C.E., 
Benson, J.M., Walker, R.C., Simon, R.S., 1990, 
in Zensus J.A.~\& Pearson, T.J., eds., Parsec-Scale
Radio Jets, Cambridge University Press, p.~152
\bibitem{wilk91} Wilkinson, P.N., Tzioumis, A.K., Benson, J.M., 
Walker, R.C., Simon, R.S., Kahn, F.D., 1991, Nature, 352, 313
\bibitem{woan} Woan, G., 1992, A\&A, 254, 25
\bibitem{worr} Worrall, D.M., 1989, in AGN and the X-ray Background,
Proc.~23rd ESLAB Symposium, Noordwijk: ESA SP-296, p.~719
\bibitem{wb} Worrall, D.M., Birkinshaw, M., 2001, ApJ, 551, 178
\bibitem{ww90} Worrall, D.M., Wilkes, B.J., 1990, ApJ, 360, 396
\bibitem{wgtz}Worrall, D.M., Giommi, P., Tananbaum, H., Zamorani, G.,
1987, ApJ. 313. 596
\bibitem{wlpr} Worrall, D.M., Lawrence, C.R., Pearson, T.J.,
Readhead, A.C.S. 1994, ApJ, 420, L17
\bibitem{wbh} Worrall, D.M., Birkinshaw, M., Hardcastle, M.J., 2001,
MNRAS,  326, L7
\bibitem{yee} Yee, H.K.C., Green, R.F., 1987, ApJ, 319, 28
\bibitem{zam} Zamorani, G., 1984, in Fanti, R., Kellermann, K., Setti,
G., eds, IAU Symposium 110, VLBI and compact radio sources, Reidel,
Dordrecht, p.~85
\end{thebibliography}
\end{document}